\documentclass{article}

\newcommand{\ie}{\mbox{i.\,e.\,\ }}
\newcommand{\iec}{\mbox{i.\,e.\,}}

\newcommand{\egc}{\mbox{e.\,g.\,}}
\newcommand{\etc}{etc.\,\ }

\newcommand{\be}{\begin{equation}}
\newcommand{\ee}{\end{equation}}

\newcommand{\ket}[1]{\ensuremath{\left|  #1 \right\rangle}}
\newcommand{\tpk}[2]{\ensuremath{\ket{#1}\!\otimes\!\ket{#2}}}
\newcommand{\twk}[2]{\ket{\begin{array}{c}\mbox{#1} \\ \mbox{#2}
\end{array}}}

\usepackage{relsize}
\usepackage{chicago}

\makeatletter
\renewcommand{\section}{\@startsection
   {section}%
   {1}%
   {0mm}%
   {-\baselineskip}%
   {0.5\baselineskip}%
   {\bfseries\normalsize\centering}}%
\makeatother

\begin{document}

\begin{center}
\LARGE
\textbf{Everett and Structure}

\vspace{0.3cm}

\textbf{\textit{David Wallace$^*$}}

\begin{figure*}[b]
(\textit{January 8, 2002})

* Centre for Quantum Computation, The Clarendon Laboratory, 
University of Oxford, Parks Road, Oxford OX1 3PU, U.K.

(\textit{e-mail:} david.wallace@merton.ox.ac.uk).

\end{figure*}

\normalsize
\end{center}

\vspace{0.7cm}

\begin{quote}
I address the problem of indefiniteness in  quantum 
mechanics: the problem that the theory, without changes to its
formalism, seems to predict that macroscopic quantities have no definite
values.  The Everett interpretation is often criticised along these lines and I shall argue 
that much of this criticism rests on a false dichotomy: that
the macroworld must either be written directly into the formalism or be
regarded as somehow illusory.  By means of analogy with other areas of
physics, I develop the view that the macroworld is instead to be
understood in terms of certain structures and patterns which emerge from
quantum theory (given appropriate dynamics, in particular decoherence).
I extend this view to the observer, and in doing so make contact with
functionalist theories of mind.

\end{quote}

\vspace{0.4cm}

\section{The measurement problem}

A simple way to think about the quantum measurement problem is as
follows:
\begin{enumerate}
\item The formalism of quantum mechanics describes the evolution of a
mathematical object called the wave-function.  By analogy with classical
physics, the natural move is to treat this wave-function as directly representing a real
thing, making it analogous to the phase-space point representing a set of particles, or to 
the vector field representing a state of the electromagnetic
field.  (The alternative of treating the wave-function as some sort of
probability distribution --- analogously to classical \emph{statistical}
mechanics --- turns out to be untenable, at least without further modification
of the theory.\footnote{There is a `statistical' or `ensemble' interpretation of quantum mechanics,
discussed by (for instance) \citeN{ballentine} and \citeN{taylor}, which does attempt to take the wave-function as 
just giving the statistical distribution of  outcomes from a large number of measurements; I find it
difficult to see how this interpretation manages to avoid both commitment to some unknown hidden-variables
theory on the one-hand, or outright anti-realism on the other, but this is not the place for such a
debate.})
\item Taking this view of the wave-function when it is used to describe
microscopic objects like atoms or molecules leads us to the conclusion
that these objects often do not have definite values of properties ---
such as spin or position --- which classically we would expect to be
definite.  This `superposition' of properties implies a very weird view 
of the microworld, but since
that world is not directly observable such weirdness is not (yet) a
problem.\footnote{At least, there is no epistemic problem; however, it might be
argued that --- pending an understanding of what (for instance) indefiniteness
of position actually \emph{means}--- our theory is simply incoherent as
a physical theory.  This suggests, as argued recently by Tappenden 
\citeyear{tappenden,tappendenforthcoming}, that we may need 
to introduce ``many-worlds'' talk at the microphysical level, before any consideration 
of macroscopic ontology.  For my own attempt to develop this approach
without having to change the quantum-mechanical formalism, see
\citeN{wallace}.}
\item However, the detailed dynamics of the quantum wavefunction
(specifically, linearity and entanglement) imply that this microscopic
indefiniteness inevitably leads to indefiniteness at the everyday level
--- so that pointers sometimes do not have definite positions, and cats
sometimes are not definitely alive or dead.  This is not merely
``weird'' but apparently pathological.
\end{enumerate}

At first sight, the obvious move seems to be to modify the theory
itself: to change either the dynamics, or the assumption that reality is
fully represented by the wave-function.  Everett's contribution to the debate 
was to challenge this `obvious' strategy and to take seriously the idea
of superpositions at the macroscopic level.  The gain of doing so would
be significant: the simple and elegant mathematical structure of quantum theory would be
left intact; there would be no need to postulate ad hoc modifications of
the dynamics, no need to add extra elements to the theory which play no 
part in its practical applications, and no conflict with relativity.

But Everett's strategy must obviously overcome major problems.  The
idea of an indeterminate macroworld seems either meaningless or just
plain contradicted by observations: what could it mean to say objects
have indefinite position?  And even if it does mean anything, surely you only
have to look at them to see that their positions are definite?

The goal of this paper is to show how these problems can be resolved,
without compromising the mathematical structure of quantum theory.  The
approach which I shall advocate is based upon decoherence theory, and
very much upon the lines of the recent versions of the Everett
interpretation proposed by \citeN{gellmann}, \citeN{saunders},
\citeN{zurekprob}, and others; in section \ref{section2} I contrast this sort of
approach to Everett with earlier versions which modify the mathematical
formalism or introduce an explicit role for consciousness.  In section
\ref{section3} I shall argue that the conceptual criticisms of the decoherence-based approach
(I don't discuss the more technical objections) are based upon a false
dichotomy (that either the macroscopic world is written directly into
the quantum formalism or it is simply an illusion), and in section
\ref{ontology} I shall defend a view of macroscopic objects which avoids
this dichotomy, based on work by Dennett (primarily
\citeNP{realpatterns}).  In sections \ref{cat}--\ref{observer} I apply this view 
to quantum mechanics, first to Schr\"{o}dinger's cat and then, in section 
\ref{observer}, to human observers.  In the latter section
I will make contact with the functionalist program 
in philosophy of mind, which fits very naturally into my framework; at the end of the section 
I briefly discuss the problem of probability in Everett interpretations, although 
for the most part I treat probability as a separate foundational problem lying largely 
outside the scope of this paper. 

\section{Recovering macroscopic definiteness}\label{section2}

Traditionally there have been two approaches taken to avoiding the
problems of macroscopic indefiniteness mentioned above, whilst preserving
the attractive features of Everett's strategy; these are now usually referred 
to as the ``Many
Worlds'' and ``Many Minds'' interpretations.  Both approaches begin with
some superposition like

\begin{eqnarray}\label{indef}
\frac{1}{\sqrt{2}}
\left(
\twk{atom}{decayed}\otimes
\! \twk{counter}{triggered}\otimes
\! \twk{observer}{detects decay}
+
\right.
\\
\nonumber
\\
\nonumber
\left.
\twk{atom}{undecayed}\otimes
\! \twk{counter not}{triggered}\otimes
\! \twk{observer detects}{no decay}
\right)
\end{eqnarray}

which \textit{prima facie} is an indefinite state in which neither the
detection apparatus nor the observer are definite.  The many-worlds
strategy interprets the two terms in this superposition as representing
two (or possibly two families of) distinct macroscopic worlds --- hence
the universal state represents a multiplicity of worlds, each one of
which is macroscopically definite.

The many-minds strategy, on the other hand, accepts that (\ref{indef})
is indefinite, and attempts to recover not definiteness but just the
\emph{appearance} of definiteness.  This is done by associating
different mental phenomena to each of the ``observer'' terms in
(\ref{indef}), so that associated with each (macroscopically indefinite) 
brain is a large number of definite minds.  Each mind sees one term in
the superposition, so that to the minds the world \emph{appears}
definite even though it is not.

However, in both of these approaches, it seems that we have to add
something to the underlying theory.  In the many-worlds case we seem to
have to specify a particular Hilbert-space basis (the so-called ``preferred 
basis'') to define worlds, and
to explain why the wave-function is to be decomposed in one way rather than
another.  Also, if the world-decomposition is defined in terms of a
basis then there would seem to be no fact of the matter as to which
world at time $t_2$ is identical to (or the successor of, etc.) a given
world at time $t_1$.  This creates pressure to add another piece of
structure, some sort of ``connection rule'' linking up worlds across
time.\footnote{This problem is discussed by, for example, \citeN{butterfield}; see also \citeN{barrett}, 
from whom I borrow the ``connection rule'' terminology.}

Arguably (and controversially! --- see \citeN{lockwood} for a defence) 
many-minds theories avoid the need to add a preferred basis to
the quantum formalism, but they do so at the price of requiring a very
close connection between fundamental physics and the philosophy of mind
--- effectively transferring the problem of selecting a basis onto our
theory of mind and requiring that theory to be explicitly 
quantum-mechanical.  The requirement for a ``connection rule'' to handle
transtemporal identity seems just as strong for many-minds theories as
for many-worlds theories: how are we to link up definite experiences at 
time $t_1$ with those at time $t_2$?

Theories can be constructed which provide this extra structure (a number
are discussed in \citeNP{barrett}) but the additions to the
formalism seem to count against the very reasons which led us to
consider Everett's strategy in the first place: the new structure is ad
hoc in the sense that it is usually quite underdetermined by observable
data, and almost inevitably spoils the relativistic covariance of the
theory.

From the 1980s onwards, decoherence theory has often been cited as part
of the solution to this problem of definiteness.  The technical details
of this approach shall not concern us here, but the basic idea is that
dynamical processes cause a preferred basis to emerge rather than having
to be specified a priori --- here we can understand `emerge' in the
sense that interference between processes described by separate terms of
the preferred basis is negligible.  (See \citeNP{zurek} for details.)

Two sorts of objection can be raised against the decoherence approach to
definiteness.  The first is purely technical: will decoherence really
lead to a preferred basis in physically realistic situations, and will
that preferred basis be one in which macroscopic objects have at least
approximate definiteness?  Evaluating the progress made in establishing
this would be beyond the scope of this paper, but there is good reason
to be optimistic.

The other sort of objection is more conceptual in nature: it is the claim
that even if the technical success of the decoherence program is
assumed, it will not be enough to solve the problem of indefiniteness.
This is because the decoherence process is only approximate: the
preferred basis is very accurately specified but not given exactly, 
and the interference between terms, though very small, is not zero.  
Furthermore, for this reason the program does not apparently help with the problem of
giving an \emph{exact} criterion for transtemporal identity.

It is this second, conceptual, objection that I wish to address in the remainder of
this paper.

\section{The fallacy of exactness}\label{section3}

The objection above arises from a view implicit in much discussion of
Everett-style interpretations: that certain concepts and objects in
quantum mechanics  must either enter the theory formally in its
axiomatic structure, or be regarded as illusions.  Consider, for
instance, Kent's influential \citeyear{kent} critique of Many-Worlds interpretations:
\begin{quote}
It's certainly true that phase information loss is a dynamical process
which needs no axiomatic formulation.  However, this is irrelevant to
our very simple point: no preferred basis can arise, from the dynamics
or from anything else, unless some basis selection rule is given.  Of
course, [Many-Worlds Interpretation] proponents can attempt to frame
such a rule in terms of a dynamical quantity - for example, some measure
of phase information loss.  But an explicit, precise rule is needed. (p.11; page
numbering refers to the internet version.)
\end{quote}
In other words, a preferred basis must either be written into the
quantum-mechanical axioms, or no such basis can exist --- the idea of
some approximate, emergent preferred basis is not acceptable. The paper goes on to make
a similar point about `worlds':
\begin{quote}
...one can perhaps intuitively view the corresponding components [of the
wave function] as describing a pair of independent worlds.  But this
intuitive interpretation goes beyond what the axioms justify: the axioms
say nothing about the existence of multiple physical worlds
corresponding to wave function components. (p.11)
\end{quote}
Analogous objections are raised about transtemporal identity:
Barrett's recent \citeyear{barrett} book gives an example.
\begin{quote}
In so far as one lacks a notion of the identity of a world over time
(and thus, no notion of the identity of an observer over time), the
splitting-worlds theory is thus empirically incoherent... But if one
adds a connection rule to the theory, then this further (because one
also needs to chose a preferred basis) detracts from the theory's
simplicity. (p.162)
\end{quote}
Barrett's quote implies that we face the same dichotomy: either there is some
precise truth about transtemporal identity which must written into the
basic formalism of quantum mechanics, or there are simply no facts at
all about the past of a given world, or a given observer.  (This seems
to be what motivates \citeN{bell}  to say that in the Everett
interpretation the past is an illusion.)

I will argue that in defending any worthwhile version of the Everett
interpretation, we should reject this view.  My claim is instead that
the emergence of a classical world from quantum mechanics is to be
understood in terms of the emergence from the theory of certain sorts of
structures and patterns, and that this means that we have no need (as well as no hope!)
of the precision which Kent and others here demand.

Before developing this account, I shall briefly address what might
appear to be a looming threat to any such approach.  The problem of
macroscopic indefiniteness is (in part) how we can
understand the quantum state as simultaneously describing two macro-objects (A and B,
say) with contradictory properties (such as being an alive cat, versus
being a dead one).  Introducing `many worlds' at the level of formalism, for all its 
disadvantages, certainly solves this problem, for then A and B are simply distinct objects. 
If however, we adopt any account in which A and B each supervene on
properties of the micro-world's ontology (say, P and Q), then if A and B have
contradictory properties then surely P and Q must themselves \emph{be} contradictory, 
and to avoid incoherence we appear to be forced back onto the explicit introduction
of `many worlds' at the level of the micro-ontology.  

There is a flaw in this argument, however.  If A and B have
contradictory properties then P and Q must certainly be
\emph{different} properties, but it does not follow that they should have
to be contradictory.  The underlying micro-ontology is (faithfully
represented by) the quantum state, and that state has a far richer set
of properties than any classical state (as can be seen, for instance,
from a position-basis viewpoint, where the quantum state of the Universe is represented
as a function over an enormously high-dimensional configuration space,
rather than the paltry three dimensions over which any classical field
is defined).  If A and B are to be `live cat' and `dead cat' then P and
Q will be described by statements about the state vector which (expressed in a position basis) 
will concern the wave-function's amplitude in vastly separated regions
$R_P$ and $R_Q$ of configuration space, and there will be no
contradiction between these statements.

\section{Understanding higher-order ontology}\label{ontology}

To see why it is reasonable to reject the dichotomy of the previous
section, consider that in science there are many examples of objepts
which are certainly real, but which are not directly represented in the
axioms.  A dramatic example of such an object is the tiger: tigers 
are unquestionably real in any reasonable sense of the word, but they
are certainly not part of the basic ontology of any physical theory.  A
tiger, instead, is to be understood as a pattern or structure in the
physical state.

To see how this works in practice, consider how we could go about
studying, say, tiger hunting patterns.  In principle --- but only in principle ---
the most reliable way to make predictions about these would be in terms of atoms and
electrons, applying molecular dynamics directly to the swirl of
molecules which make up tigers and their environment.  In practice,
however, this is clearly insane: no remotely imaginable computer would
be able to solve the $10^{35}$ or so simultaneous dynamical equations
which would be needed to predict what the tigers would do, and even 
if such a computer could exist its calculations could not remotely 
be said to \emph{explain} their behaviour.

A more effective strategy can be found by studying the structures 
observable at the multi-trillion-molecule level of
description of this `swirl of molecules'.  At this level, we will
observe robust --- though not $100\%$ reliable --- regularities, which
will give us an alternative description of the tiger in a language of
cells and molecules.  The principles by which these cells and molecules
interact will be derivable from the underlying microphysics, and will
involve various assumptions and approximations; hence very occasionally 
they will be found to fail.  Nonetheless, this slight riskiness in our
description is overwhelmingly worthwhile given the enormous gain in
usefulness of this new description: the language of  cell biology is
both explanatorily far more powerful, and practically far more useful,
than the language of physics for describing tiger behaviour.

Nonetheless it is still ludicrously hard work to study tigers in this
way.  To reach a really practical level of description, we again look
for patterns and regularities, this time in the behaviour of the cells
that make up individual tigers (and other living creatures which
interact with them).  In doing so we will reach yet another language,
that of zoology and evolutionary adaptationism, which describes the
system in terms of tigers, deer, grass, camouflage and so on.  This 
language is, of course, the norm in studying tiger hunting patterns, and
another (in practice very modest) increase in the riskiness of our
description is happily accepted in exchange for another phenomenal rise
in explanatory power and practical utility.

Of course, talk of zoology is grounded in cell biology, and cell biology
in molecular physics, but we cannot discard the tools and terms of
zoology to work directly with physics, without (a) losing explanatory
power, and (b) taking forever.

What moral should we draw from this mildly fanciful example?  That
higher-level ontology is to be understood in terms of pattern or
structure: in a slogan,
\begin{quote}
\textbf{A tiger is any pattern which behaves as a tiger.}
\end{quote}

More precisely, what we have is a criterion for which patterns are to be
regarded as real, which we might call Dennett's criterion (in
recognition of a very similar view  proposed by
\citeNP{realpatterns}\footnote{A more restricted proposal of this sort (applying specifically
to intentional systems, such as the chess computer example given here) was made by Dennett 
significantly earlier, in \citeN{intentionalsystems}.  Though the more general view is implicit
in many of Dennett's earlier writings it does not seem to have been states explicitly prior to
the \citeyear{realpatterns} paper I cite.}).
\begin{quote}
\textbf{Dennett's Criterion:} A macro-object is a pattern, and the existence 
of a pattern as a real thing
depends on the usefulness --- in particular, the explanatory power and predictive 
reliability --- of theories which admit
that pattern in their ontology.
\end{quote}
Dennett's own favourite example is worth describing briefly in order to
show the ubiquity of this way of thinking: if I have a computer running
a chess program, I can in principle predict its next move by analysing
the electrical flow through its circuitry, but I have no
chance of doing this in practice, and anyway it will give me virtually
no understanding of that move.  I can achieve a vastly more effective
method of predictions if I know the program and am prepared to take the
(very small) risk that it is not being correctly implemented by the
computer, but even this method will be practically very difficult to
use.  One more vast improvement can be gained if I don't concern myself
with the details of the program, but simply assume that whatever they
are, they cause the computer to play good chess.  Thus I move
successively from a language of electrons and silicon chips, through one
of program steps, to one of intentions, beliefs, plans and so forth ---
each time trading a small increase in risk for an enormous increase in
predictive and explanatory power.\footnote{It is, of course, highly
contentious to suppose that a chess-playing computer \emph{really}
believes, plans \etc  Dennett himself would embrace such claims (see
\citeN{intentional} for an extensive discussion), and they are at least
suggested by the functionalist program in philosophy of mind which I
discuss in section \ref{observer}.  However, for the purposes of this section there
is no need to resolve the issue: the computer can be taken only to
`pseudo-plan', `pseudo-believe' and so on, without reducing the
explanatory importance of a description in such terms.}

Why is it reasonable to claim, in examples like these, that higher-level 
descriptions are \emph{explanatorily} more powerful than lower-level ones? 
In other words, granted that a prediction from microphysics is in
practice impossible, if we had such a prediction why wouldn't it count
as a good explanation?  To some extent I'm inclined to say that this is
just obvious --- anyone who really believes that a description of the trajectories followed
by the molecular constituents of a tiger explains why that tiger eats a deer means
something very different by `explanation'.  But possibly a more
satisfying reason is that the higher-level theory to some extent `floats
free' of the lower-level one, in the sense that it doesn't care how its
patterns are instantiated provided that they are instantiated.   (Hence
a zoological account of tigers requires us to assume that they are
carnivorous, have certain strengths and weaknesses, and so on, but
doesn't care what their internal makeup is.)  So an explanation in terms
of the lower-level theory contains an enormous amount of extraneous
noise which is irrelevant to a description in terms of higher-level
patterns.  See \citeN{putnam} for further description of this point.

This approach to higher-order ontology applies to physics itself as well
as to theories other than physics, as illustrated by one further example:
that of quasi-particles.  To understand these, consider vibrations in a
(quantum-mechanical) crystal.  These can in principle be described
entirely in terms of the individual crystal atoms and their quantum
entanglement with one another --- but it turns out to be overwhelmingly
more useful to think in terms of `phonons' \ie collective excitations of
the crystal which behave like `real' particles in most respects.  

This sort of thing is ubiquitous in solid-state physics, and the
collective excitations are called `quasi-particles' --- so crystal vibrations are
described in terms of phonons, waves in the
magnetisation direction of a ferromagnet in terms of
magnons, collective electron waves in a plasma in terms of plasmons, and
so on.  But are quasi-particles real?  Well, they can be created and
annihilated; they can be detected (by, for instance, scattering them off
`real' particles like neutrons); in some cases (such as so-called `ballistic' phonons) 
their time-of-flight can be measured; and they play a
crucial explanatory role in solid-state theories.\footnote{Any solid-state textbook is replete with
explanations of empirical phenomena which are couched in terms of quasi-particles; 
see \citeN{kittel}, for instance.}  We have no more
evidence than this that `real' particles exist, so it seems absurd to
deny the existence of the quasi-particles.

But when \emph{exactly}, you might ask, are quasi-particles present?  This question
has no precise answer.  It is essential in a quasi-particle formulation
of a solid-state problem\footnote{See the first chapter of \citeN{abrikosov} 
for a discussion.} that the quasi-particles decay only slowly
relative to other relevant timescales (such as their time-of-flight)
and when this criterion (and similar ones) are easily met then quasi-particles are
definitely present.  When the decay rate is much too high, the quasi-particles 
decay too rapidly to behave in any `particulate' way, and the
description becomes useless: hence we conclude that no quasi-particles
are present.  However, clearly it is a mistake to ask \emph{exactly}
when the decay time is short enough ($2.54\;\times$ the interaction time?)
for quasi-particles not to be present.  What actually happens is that,
as we lower the decay time, the quasi-particle description becomes less
and less advantageous compared to a lower-level description in terms of
crystal atoms --- hence by Dennett's criterion it becomes less and less
viable to regard them as real, until ultimately they are clearly no
longer of any use in studying the crystal and we must either revert to the 
underlying description or look for another, more useful higher-level
distinction.  But the somewhat blurred borderline
between states where quasi-particles exist and states where they don't
should not undermine the status of the quasi-particles as real --- any more
than the absence of a precise point where a valley stops and a mountain
begins should undermine the status of the mountain as real.

(In fact, although this account of quasi-particles represents them as structures in an ontology
of `real' particles, the description in
terms of nonrelativistic particle mechanics is itself effective, and
derives from a description in terms of quantum field theory --- 
there is every reason to believe particles like quarks and electrons to
be patterns in the underlying quantum field in almost exactly the same
sense that quasi-particles are patterns in the underlying crystal.  It
is interesting to ask whether the existence of \emph{some} underlying
`stuff' is essential, or whether we can continue this chain of theories
forever; such a question lies beyond the scope of this paper, though.)

This view of
higher-order ontology as pattern or structure has some consequences which, though obvious
given the nature of patterns, will play an important role in the later discussion of quantum 
mechanics.
\begin{enumerate}
\item Patterns can be imprecise.  As the quasi-particle example
should illustrate, a pattern can tolerate a certain amount of `noise' or
imprecision whilst still remaining the same pattern.  (A tiger which
loses a hair is still the same tiger).  Beyond a certain point the noise
is such that the pattern can no longer be said to be present, but there
is no reason to expect there to be any precise point where this occurs.
(It may sometimes be convenient to define such a point by fiat: the
biologist sometimes introduces an exact moment when one species becomes
another; the astrophysicist defines an exact radius at which the sun's
atmosphere starts.  But neither believes that any deep truth is captured
by this exactness.)
\item Patterns may involve dynamics, or be temporally extended.
A `pattern' in the sense I am using it need not be realised at an
instant, but may depend on the behaviour over some timescale of the
constituents of a pattern - what distinguishes a tiger from an inanimate facsimile of
one is the behaviour of the former, not its shape.
\item There is a concept of transtemporal identity for patterns, but
again it is only approximate.  To say that a pattern $P_2$ at time $t_2$ is
the same pattern as some pattern $P_1$ at time $t_1$ is to say something
like ``$P_2$ is causally determined largely by $P_1$ and there is a
continuous sequence of gradually changing patterns between them'' ---
but this concept will not be fundamental or exact and may sometimes
break down.  
\end{enumerate}

Before ending this section, I should acknowledge that my account is obviously linked to the
topic of how one theory can emerge from, or be reduced to, another --- and that this latter topic is
highly controversial.  Space does not permit any detailed engagement with the extensive literature
on the subject, but I give here a few recent references: \citeN{butterfieldisham} give
a general discussion of emergence using time in quantum gravity as an example; \citeN{thalos} discusses
the tension between physics and `higher-level' sciences, in the context of social science, and 
\citeN{auyang} is concerned with the way in which complex behaviour 
emerges from the interaction of simple systems; she uses quasi-particles as an example, in fact.
There is also some overlap with the current debate on structural realism (proposed originally by 
\citeN{worrall}, developed by, \egc, \citeN{ladyman}, and criticised by, \egc, \citeN{psillos}).

\section{Quantum theory in structural terms}\label{cat}

In order to show how the ideas of the last section apply to quantum mechanics, we consider
the time-honoured problem of Schr\"{o}dinger's cat.  Recall the
situation: our unfortunate cat is locked in a box and at some time ---
let us say noon --- an unstable atomic nucleus is measured by a device within the box.  If the
device finds the atom to be undecayed the cat lives, but if it finds that it is decayed
then poison gas is released into the box.  If the atom's state is
indefinite just before the measurement, then so is the cat's state just
after the measurement.

Now, suppose that the cat is put into the box at 11am and we are asked
to predict what happens to it in the next hour.  We do not know the
wavefunction of the cat at this point, and even if we did know it
exactly it would be of little use to us, for we cannot possibly solve
the Schr\"{o}dinger equation for such a complicated system --- nor can
we even solve some sort of classical or semiclassical approximation to
it.  

Nonetheless we can say useful things about the cat:
\begin{itemize}
\item from solid-state physics we can predict that the cat won't
spontaneously vaporize;
\item from animal physiology we can predict that the cat won't
spontaneously die or grow a second tail;
\item from cat psychology we can predict that the cat won't start eating
itself, and will probably remain asleep for the whole hour.
\end{itemize}
It is because of the power of this cat-level description to tell us
about the future evolution of the wave-function, and because of the
unavoidable need to work at cat-level in considering that future
evolution, that we say --- via Dennett's criterion --- that there is a
cat present in the system.

Now consider the evolution of the system after twelve noon, when the
measurement is made, but suppose that the atomic nucleus, instead of
being in an indefinite state, either definitely did or definitely did
not decay.  In each case, to predict the system's behaviour in the next
hour, we use exactly the same methods --- \egc, if the cylinder of
poison gas breaks, then cat psychology tells us that the cat will probably
jump backwards, and animal physiology tells us that it will die and in
due course start to decompose.

Now, quantum mechanics is linear.  If we know what happens if the atom
definitely does, or definitely does not, decay, then we can predict what
happens if we have a superposition of decaying and not decaying.
However, in doing so we are using exactly the same methods as before: we
are taking advantage of the patterns present in the two branches of the
wave-function.  In other words --- and this is the crucial point ---
\emph{in each of the branches there is a `cat'
pattern, whose salience as a real thing is secured by its crucial
explanatory and predictive role}.  Therefore, by Dennett's criterion
there is a cat present in \emph{both} branches after measurement.\footnote{If there 
are two cats, didn't the mass of the universe just increase?  It is easy to see mathematically
that this is not the case: the mass of the universe is a property of the universal state, given by
the expectation value of some `mass operator' relative to that state.  As such, although we can
quite happily talk about mass within a given branch, mass is simply not additive across
branches --- any more than a superposition of two states each with energy $E$ does not have energy $2E$. 
Analogously, a cat may have mass $m$ in the morning and mass $m$ in the afternoon but the cat (regarded
as a temporally extended object) still only has mass $m$, not $m+m$.  (I am grateful to Simon Saunders
for this analogy).}

Is it the same cat?  Well, it is a future version of the same cat, in
the sense described in the previous section: \iec, it is a pattern
causally determined by the original cat and linked to it by a
continuously changing sequence of cat patterns.  It's really just a
matter of terminology 
whether we decide that the whole branching set of
living and dead cats `is the same cat' (as defended in \citeNP{tappenden}); the 
point to be learned, though, is that when describing patterns
we shouldn't expect any more from transtemporal identity than
approximate, `effective' concepts which sometimes break down.  (See \citeN{wallace} for 
further discussion of identity over time in quantum mechanics.)

Another question which at first sight should have a precise answer: if
there was one cat before the measurement and two after it, when
\emph{exactly} did the duplication of cats occur?  But first sight is
mistaken.  Before the decay there is certainly one cat.  When the
measurement occurs we will have a coherent superposition of both
measurement outcomes --- but after a very short time decoherence will
remove the interference between these branches, and after this time
there will be two cats present.  During the decoherence period the
wavefunction is best regarded as some sort of `quantum soup' which does
not lend itself to a classical description --- but since the decoherence
timescale $\tau_D$ is incredibly short compared to any timescale
relevant at the cat level of description, this need not worry us.  Put another way, the cat
description is only useful when answering questions on timescales far
longer than $\tau_D$, so \emph{whether or not} quantum splitting is occurring,
it just doesn't make sense to ask questions about cats that depend on
such short timescales.

\section{Superpositions of patterns}\label{systemfallacy}

To see in a different way how the ideas of Sections \ref{ontology}-\ref{cat}
resolve the problem of
macroscopic indefiniteness, consider the following sketch of the problem.
\begin{enumerate}
\item After the experiment, there is a linear superposition of a live
cat and a dead cat.
\item Therefore, after the experiment the cat is in a linear
superposition of being alive and being dead.
\item Therefore, the macroscopic state of the cat is indefinite.
\item This is either meaningless or refuted by experiment.
\end{enumerate}
But (1) does not imply (2).  The belief that it does is based upon an
oversimplified view of the quantum formalism, in which there is a
Hilbert space of cat states such that any vector in the space is a
possible state of the cat.  This is superficially plausible in view of
the way that we treat microscopic subsystems: an electron or proton, for
instance, is certainly understood this way, and any superposition of
electron states is another electron state.  

But any state of a cat is actually a member of a Hilbert space
containing states representing all possible macroscopic objects made out
of the cat's subatomic constituents.  Because of Dennett's criterion, this 
includes states which describe 
\begin{itemize}
\item a live cat;
\item a dead cat;
\item a dead dog;
\item this paper \ldots
\end{itemize}

We can say (if we want, and within nonrelativistic quantum mechanics\footnote{The 
situation is much more complicated in quantum field
theory, where the particles emerge in an effective way themselves and
where particle number is not conserved.    The only exactly definable
subsystem would be the field degrees of freedom in the spatial vicinity of the
cat, but this fails to allow for the fact that the cat might move out of that region.  
In QFT even more than NRQM we are forced to a pattern concept of macroscopic systems.})
that the particles which used to make up the cat are now jointly in a
linear superposition of being a live cat and being a dead cat.  But cats
themselves are not the sort of things which can be in superpositions.
Cats are by definition ``patterns which behave like cats'', and there
are definitely two such patterns in the superposition.  

The point can be made more generally: 

\textbf{It makes sense to consider a
superposition \emph{of} patterns, but it is just meaningless to speak of
a given pattern as being \emph{in} a superposition. } 

Thus, a pattern
view of macroscopic ontology essentially solves the problem of
indefiniteness by replacing indefiniteness with multiplicity; since it does so at the level of macroscopic
objects including inanimate ones, it is closer in spirit to a Many-Worlds approach than to a 
Many-Minds one.  However, this multiplication of patterns happens naturally within the
existing formalism, and does not need to be added explicitly to the
formalism.

It is important to remain clear what macro-objects are patterns
\emph{in}: they are not patterns in the positions of micro-objects, or
in fundamental fields; they are patterns in (the properties of) the
quantum state.  As mentioned in the footnote on page 2, we can and do
remain neutral about how this state is itself to be interpreted, since
all we need from it are its structural properties, such as: what its
representation is in the eigenbasis of a given operator.  Of course,
without specification of some set of preferred operators the state is
structureless (we can say that it is a vector of unit norm in a
countably-infinite-dimensional complex Hilbert space, but that's about
it).  The details of this specification depend on the particular quantum
theory with which we are working: in non-relativistic quantum mechanics,
for instance, they are given by the generators of the Gallilei group for
individual particles, whilst in quantum field theory they are
given in terms of the map between spacetime regions and operator algebras 
(see \citeN[section 2.2]{wallaceqft} for a discussion).  

As an aside, the  analysis of this paper gives support to Deutsch's 
claim that the de Broglie-Bohm pilot-wave theory\ \cite{bohm,holland} and its variants are
``parallel-universes theories in a state of chronic
denial''\cite{deutschbjps}.  In such theories\footnote{I confine my attention here
to those versions of the pilot-wave theory in which the wave-function is taken to be
physically real.  Those versions in which only the corpuscles exist are, in my view, 
in even greater denial, but I shall not argue the point here.  (See \citeANP{wallacedarg} (forthcoming) for a discussion).}
the wave-function is supplemented by a collection of `corpuscles',
particles guided by the wave-function and supposed to define our
observed universe.  But to predict the behaviour of the corpuscles we
have to predict the behaviour of the wave-function, and to predict the
behaviour of the wave-function we have to study the emergent patterns
within it.  Thus cats and all other macro-objects can be identified in the structure
of the wave-function just as in the structure of the corpuscles.  But
the patterns which define them are present even in those parts of the
wave-function which are very remote from the corpuscles.  So if we
accept a structural characterisation of macroscopic reality, we must
accept the multiplicity of that reality in the de Broglie-Bohm pilot
wave as much as in the Everettian universal state.  

\section{The role of the observer}\label{observer}

We have not yet considered explicitly how observers are to fit into the
framework just described.  However, if we are happy to extend Dennett's
criterion to conscious observers, then they fit into the framework quite
straightforwardly: if a tiger is any pattern which behaves like a tiger,
then an observer is any pattern which behaves like an observer.  

This is
essentially an expression of an established viewpoint in the philosophy
of mind: functionalism.  Though there are many versions of
functionalism, for our purposes we can define it as follows.
\begin{quote}
\textbf{The functionalist claim}: As a matter of
conceptual necessity,\footnote{The phrase `conceptual necessity' is
supposed to indicate that, according to functionalism, mental properties are to be 
understood as by definition being present in systems with certain
functional properties.  We can distinguish this from the weaker claim
that, \emph{in fact}, mental properties are present in a system if and only if
that system has certain functional properties; much of what I will say
goes through in this case also.  See \citeN{chalmers} for a recent presentation of
this weaker claim.}
mental properties are supervenient on structural
and functional properties of physical systems, and on no other
properties.  Hence, it doesn't matter what a brain is made of, only how
it works.
\end{quote}
Functionalism is at the root of the artificial intelligence project, for it entails that
any sufficiently accurate computer simulation of a conscious being will
itself be conscious.  I will not attempt to defend it here, but will simply explore 
its implications for quantum theory.\footnote{See
\citeN{lewis} and \citeN{armstrong} for two classic defences of functionalism; 
\citeN{dennettconsciousness} for a more recent defence;  
\citeN{block}, \citeN{chalmers}, \citeN{penroseenm} and \citeN{searle} for
a variety of criticisms; and \citeN{hofstadterdennett} for a collection of articles against and
(mostly) for functionalism.}  

Given functionalism, we can see that quantum mechanics implies the
multiplication of observers in just the same way as it does the
multiplicity of cats.  To see this in rather more detail, let us consider an idealised
measurement of some 2-state system: the system is assumed to be measured
in some basis $(\ket{1},\ket{2})$.  First consider the case where the 
2-state system is actually in state \ket{1}, then the observer's state
will remain definite after the measurement.  Let's suppose the joint
state of 2-state system and observer some time $t$ after the measurement 
is 
\be \ket{\psi_t;1} = \tpk{1}{f_1(t)},\ee
where $f_1(t)$ is some functional process describing the observer in the
time following his observation of \ket{1}, and \ket{f_1(t)} 
(for varying $t$) is the sequence of states realising that process.
Similarly if the 2-state system is actually in state \ket{2}, the joint
state post-measurement will be
\be \ket{\psi_t;2} = \tpk{2}{f_2(t)}.\ee
In accordance with comment (2) above, the states \ket{f_1(t)} and \ket{f_2(t)} 
describe not just the observer,
but an entire macroscopic region (where objects in that region are
defined in structural terms, as explained above).

Now let the 2-state system be in some superposition $\alpha \ket{1} +
\beta \ket{2}$.  Linearity tells us that the overall state at time $t$
must be
\be \ket{\psi_t} = \alpha \tpk{1}{f_1(t)} + \beta \tpk{2}{f_2(t)}.\ee
Now, each process $f_1(t)$,
$f_2(t)$ describes (the mind of) an observer.  The state \ket{\psi_t} realises both such processes;
 hence, it describes two observers (rather than one observer in
an indefinite state --- whatever that might mean), and we have again
replaced superposition by multiplicity.\footnote{Note that my argument is rather different 
from that used by \citeN{chalmers} to take superposition into multiplicity.  Chalmers proposes 
a principle (the `superposition principle') which effectively says that if conscious experience
is present in one term of a superposition, then it is present in the superposition; this was shown
to be unworkable by \citeN{byrnehall}.  I make use only of the much weaker result (following from 
the functionalist criterion) that a superposition
of orthogonal states, each of which is determinately part of a sequence of functional states, 
realises all the functional processes encoded by those sequences.  This in turn relies upon the
existence of decoherence to give a preferred basis in which functional sequences are possible;
in this use of a preferred basis my approach is similar to that suggested by \citeN{vaidmanreply}
in his reply to Byrne and Hall.}

We can see, then, that worries about observers with indefinite mental
states are as misplaced as worries about cats which are indefinitely
alive or dead. Patterns are not superposed, but duplicated, by
the measurement event, and ultimately we are regarding mental
states as just special sorts of patterns (although these
patterns need not be \emph{visual} patterns, realised in the
instantaneous physical state; rather, they are likely to be
\emph{behavioural} patterns, which describe regularities in the dynamics
of the physical state as well as its instantaneous configuration --- c.\,f.\, \citeNP{realpatterns}.)

We can also consider how our observer views the measurement event.  In
the cases where the state of the system being observed had been
definitely either \ket{1} or \ket{2}, the observer's pre-measurement
process ($f_0(t)$, let us say) would have changed unproblematically
into $f_1(t)$ or $f_2(t)$, and the observer would certainly interpret
this as personal survival: hence $f_0(t)$ and $f_i(t)$ describe the same
person.  It is then legitimate for the observer to understand the
measurement as himself surviving as two diverging copies (of different
weights) following the measurement.  (As \citeN{saunders} has pointed
out, this is closely analogous to the cases of personal fission
considered by \citeN{parfit}.)

(It is tempting to ask: What does it feel like while the split itself is occurring?  Hopefully it should be clear by
now that this is a bad question: if (as functionalism claims) statements
about mental phenomena are statements about the functional behaviour of
the brain (\iec, about the dynamical patterns in it) and if the
timescales on which the functional processes occur are very long compared
to the decoherence timescale (which they are) then there can be no
awareness of the event of splitting at all --- thus allowing us to justify Everett's
famous claim to this effect (Everett \citeyearNP{everett}, p.\,460).  By analogy, suppose an
artificial-intelligence program were to be run on a (classical) digital
computer; it would be meaningless to ask what it felt like for that
program whilst the computer was in the process of changing from one
digital configuration to another.  Understanding that process requires us
to abandon the language of computer programs and descend to the level of
electronics, and `mental' talk about the computer or program doesn't
engage with that level.)

The approach advocated here also alleviates (though does not solve entirely) the problem of probability
in the Everett approach.  An observer about to measure a superposed state 
knows that after the measurement there will exist more than one
functional structure which he will regard as the same individual as
himself.  He has no reason not to care about their futures just as he
cares about `his own' future, for even in the absence of splitting his
future existence consists only in the future presence of patterns such
as these.  But the different future copies may have different interests
which he could influence by actions prior to the measurement; so how are
these different interests to be weighted?  There is no a priori reason
to weight them equally.  Granted, we have not shown that the `correct'
or `most rational' weighting is the standard one, but we have at least
shown that it is rational for the observer to assign some weighting: in
other words, we have shown that there is room for probabilistic concepts (at least the
decision-theoretic sort) to be accommodated in the theory.  This is
already enough to bring the Everett interpretation onto the same level as
any other physical theory, for --- as pointed
out in the quantum context by \citeN{papineau} --- we have no really
satisfactory understanding of probability in any other context either!
For more constructive attempts to justify the probability rules, though
(from a wide variety of perspectives) see \citeN{deutschprob},
 \citeN{saunders}, \citeN{tappenden}, \citeN{vaidman} and \citeN{zurekprob}.

It is worth remembering the crucial role that  decoherence is playing 
in this account: without
it, we would not have the sort of branching structure which allows the
existence of effectively non-interacting multiple near-copies of a given
process.  As it is, though, we are able to identify many different
functional structures realised in different parts of the universal
state, each with the right sort of complexity to merit the title
`observer'.

Would it be possible to reject functionalism --- that is, reject the application of
Dennett's criterion to conscious observers --- without having to reject this paper's
`structural' approach to quantum theory?  Not necessarily, for
functionalism is neutral about how functional systems are to be realised
physically, whereas in this structural approach to quantum theory 
there is space for us to require the system to be
instantiated in a certain way --- say, in the position basis (although see \citeN{wallace} for
the difficulties of this particular basis choice).  However,
the structural approach is committed to an approach to the mind which 
\begin{itemize}
\item denies observers some uniquely special status, but describes them as emergent 
as structures and patterns in lower-level physics (specifically, in lower-level \emph{classical}
physics, itself to emerge from unitary quantum physics via decoherence);  
\item is comfortable with some rough edges in the definition of which
systems count as observers (for decoherence will never give us an exact
macroworld).
\end{itemize}
Functionalism fits these criteria in a very natural way.

\section{Conclusions}

In his critique of many-worlds interpretations, \citeN{kent} states that 
\begin{quote}
[W]e have tried to clarify the logical structure of the MWI \ldots The
attempt may not have entirely succeeded.  But we are convinced that the
\emph{procedure} is justified, and in fact that axiomatization should
have been insisted upon from the beginning.  For any MWI worth the
attention of physicists must surely be a physical theory reducible to a
few definite laws, not a philosophical position irreducibly described by
several pages of prose.
\end{quote}
It is not the purpose of this paper to argue against this view of
physical theories.  I agree that physical theories should be
axiomatizable, and in fact would say that the axioms of any worthwhile Everettian
theory should be just those of `bare' unitary quantum mechanics, without
axioms of measurement or collapse.

However, when we \emph{are} describing observations, or cats, or people,
within physics, we inevitably need to make contact with higher-order
theories --- of material science, of cat biology, of psychology and
neuroscience.  This contact is not made by fiat, via abstractly or generally stated
principles;\footnote{And, it should be stressed, Kent certainly makes no such
claim.} rather, it occurs because those theories are emergent from the microphysics, 
describing patterns which occur \emph{within} the microphysics.  Indeed we do need several
pages of somewhat philosophical prose to describe carefully how this
emergence takes place --- but the point is that having understood the
process in the classical case, there really is no reason to think
anything different is going on in quantum theory.  

To summarise the view of quantum theory that then emerges:

\begin{itemize}
\item Macroscopic objects are to be understood as structures and
patterns in the universal quantum state.
\item Multiplicity occurs at the level of structure --- thus macroscopic
objects do not have indeterminate states after quantum measurements, but
are genuinely multiplied in number.
\item We can tolerate some small amount of imprecision in the
macroworld: a slightly noisy pattern is still the same pattern.  Hence
we do not need to worry that decoherence does not give \emph{totally}
non-interfering branches, just very nearly non-interfering ones.
\item There will be no precise answers to some questions (such as, `when
did the splitting take place?'), just very accurate ones.
\item Other questions (such as those concerning transtemporal identity, or identity between
objects across branches) will
not always have good answers at all, because they rely on concepts which
though practically very useful, sometimes break down.
\end{itemize}

\small
\vspace{0.5cm}

\noindent \emph{Acknowledgements} --- Although I haven't met Daniel Dennett, this paper has been
enormously influenced by his work.  I am also very grateful to Katherine Brading, Jeremy Butterfield,
Paul Tappenden
and Lev Vaidman for comments on this paper; to Harvey Brown, Clare Horsman, Adrian Kent, 
Simon Saunders, and Wojciech Zurek for useful discussions; and to an anonymous referee
for drawing my attention to the argument presented at the end of section \ref{section3}.

\end{document}